\documentclass[11pt]{article}

\usepackage{epsf}
\usepackage{url}

\usepackage{amssymb,lineno}

\usepackage[dvips]{graphicx}
\newtheorem{prob}{Problem}
\newtheorem{alg}{Algorithm}
\newtheorem{theo}{Theorem}
\newtheorem{lem}{Lemma}
\newtheorem{clm}[lem]{Claim}

\newcommand{\BA}{\begin{alg}} \newcommand{\EA}{\end{alg}}
\newcommand{\BE}{\begin{enumerate}} \newcommand{\EE}{\end{enumerate}}
\newcommand{\BT}{\begin{theo}} \newcommand{\ET}{\end{theo}}
\newcommand{\BL}{\begin{lem}} \newcommand{\EL}{\end{lem}}
\newcommand{\BCM}{\begin{clm}} \newcommand{\ECM}{\end{clm}}
\newcommand{\BI}{\begin{itemize}} \newcommand{\EI}{\end{itemize}}

\def\FullBox{\hbox{\vrule width 8pt height 8pt depth 0pt}}
\newcommand{\qed}{\;\;\;\FullBox}
\newenvironment{prf}{\noindent{\bf Proof:~~}}{\(\qed\)}
\newcommand{\BPF}{\begin{prf}} \newcommand {\EPF}{\end{prf}}
\newenvironment{proofof}[1]{\noindent{\bf Proof of {#1}.~}}{\endprf}
\newcommand{\BPFOF}{\begin{proofof}} \newcommand {\EPFOF}{\end{proofof}}

\newcommand{\BEQN}{\begin{eqnarray}}\newcommand{\EEQN}{\end{eqnarray}}
\newcommand{\BEQ}{\begin{equation}} \newcommand{\EEQ}{\end{equation}}

            \newcommand{\eat}[1]{}

\begin{document}

\title{An $O(\log n)$-approximation for the Set Cover Problem with Set Ownership}  % how do you like this title?

\author{Mira Gonen
   and
Yuval Shavitt\\
{School of Electrical Engineering,
    Tel Aviv University,
         Ramat Aviv 69778, Israel.}
}

%\author{Mira Gonen}
%\ead{gonenmir@post.tau.ac.il}

%\author{Yuval Shavitt}
%\ead{shavitt@eng.tau.ac.il}
%\address{Tel-Aviv University,
%    Ramat Aviv, {\sc Israel}}
%\address{Tel-Aviv University,
%    Ramat Aviv, {\sc Israel}
 %   }

%\ead[url]{authors.elsevier.com/locate/latex}

%\author{Mira Gonen \\
%    Department of EE  -- Systems \\
%    Tel-Aviv University  \\
%    Ramat Aviv, {\sc Israel} \\
%    {\tt gonenmir@post.tau.ac.il}
%\and
%    Yuval Shavitt \\
%    Department of EE  -- Systems \\
%    Tel-Aviv University  \\
%    Ramat Aviv, {\sc Israel} \\
%    {\tt shavitt@eng.tau.ac.il}
%}

%\date{}

\maketitle
\begin{abstract}
In highly distributed Internet measurement systems distributed
agents periodically measure the Internet using a tool called {\tt
traceroute}, which discovers a path in the network graph. Each agent
performs many traceroute measurement to a set of destinations in the
network, and thus reveals a portion of the Internet graph as it is
seen from the agent locations. In every period we need to check
whether previously discovered edges still exist in this period,
 a process termed {\em validation}. For this end we maintain a database of
  all the different measurements
 performed by each agent.
 Our aim is to be able to {\em validate} the existence of all
 previously discovered edges in the minimum possible time.

  In this work we formulate the validation problem as a generalization of
   the well know set cover problem. We reduce the set cover problem to the validation
  problem, thus proving that the validation problem is ${\cal NP}$-hard.
  We present a $O(\log n)$-approximation algorithm to
  the validation problem, where $n$ in the number of edges that need
  to be validated. We also show that unless ${\cal P = NP}$ the approximation
  ratio of the validation problem is $\Omega(\log n)$.
\end{abstract}

%\begin{keyword}
%Internet, measurement systems, traceroute
%\end{keyword}

\section{Introduction}\label{intro.sec}
 Our problem arise in the context
of highly distributed Internet measurement systems
\cite{dimes-CCR,dimes-book}.  In this type of systems, distributed
agents periodically measure the Internet using a tool called {\tt
traceroute}, which discovers a path in the network
graph\footnote{The path can be expressed at various levels of
abstraction. The most common level in use is the autonomous system
(AS) level, where each node in the graph (and thus in the path)
represent an AS (or a network) in the Internet. }. Each agent performs
many traceroute measurement to a set of destinations in the network,
and thus reveals a portion of the Internet graph as it is seen from
the agent locations. While some edges can be seen from many
measurement locations, others can be seen only from a handful
locations \cite{dimes-CCR,dimes-book,SToC-bias}, which is the major
reason for distributing this process. We create a periodic map by
unifying the measurements made by all the agents over this period.

There are many possible heuristics to direct agents to destinations
in order to find as many graph edges as possible. However, one thing
we have to do in every period is to check whether previously
discovered edges still exist in this period,
 a process termed {\em validation}.  For this end we maintain a database of all the different measurements
 performed by each agent\footnote{The list is kept at the abstraction level
 we are interested in, e.g., at the AS level.}.
 Our aim is to be able to {\em validate} the existence of all
 previously discovered edges in the minimum possible time.

A solution to the validation problem is to model each tracroute
measurement as a set of edges, and then look for the smallest group
of traceroute measurements (the sets) that covers the known graph,
e.g., using a set cover logarithmic approximation algorithms
\cite{Set-Cover-j}. However,  this solution may end up finding many
groups which are measured by one agents while leaving other agents
with little or no measurements to perform. Since all agents measure
at roughly the same rate, the termination time of the validation
task is determined by the time it will take the agent with the
largest numbers of measurements to complete its task. Thus, our aim
is not to minimize
 the number of measurement that cover the graph, but to minimize the maximal
 number of measurement which is assigned to the agent with the most measurements.
 Therefore reducing the validation problem to the set cover problem
 will not necessarily give us the best solution, so we describe the validation problem as a
 generalization of the set cover problem.

 \textbf{Our Results.} We define a new generalization of the set
 cover problem that is equivalent to the validation problem, and
 give an $O(\log n)$-approximation algorithm, where $n$ is the number of
 edges in the validation problem, and show that our approximation ratio is tight, namely
 that our generalization of set cover cannot be approximated in polynomial time to within a factor of $o(\log n)$.

 \emph{Organization:} In Section~\ref{prelim.sec} we give notations and a formal
 definition of the problem. In Section~\ref{approx.sec} we present an $O(\log n)$-approximation
 algorithm for the generalized set cover problem, and prove that this ratio cannot be asymptotically improved.

\section{Preliminaries}\label{prelim.sec}
For an algorithm $\textbf{A}$, denote the objective value of a
solution it delivers on an input $\emph{I}$ by
$\textbf{A}(\emph{I})$. An optimal solution is denoted by
$\textsc{opt}$, and the optimal objective value is denoted by
$\textsc{opt}$ as well. The (absolute) approximation ratio of
$\textbf{A}$ is defined as the infimum $\rho$ such that for any
input $\emph{I}$, $\textbf{A}(\emph{I})\le \rho \cdot
\textsc{opt}(\emph{I})$.

Given a universe $U = \{u_1,...,u_n\}$ and a family of its subsets,
${\cal S} = \{S_1,...,S_k\}\subseteq P(U)$, $\bigcup_{S_j\in {\cal
S}}{S_j}= U$, set cover is the problem of finding a minimal
sub-family $\bar{\cal S}$ of ${\cal S}$ that covers the whole
universe, $\bigcup_{S_j\in {\bar{\cal S}}}{S_j}= U$. Set cover is a
classic ${\cal NP}$-hard combinatorial optimization problem, and it
is known it can be approximated to within $\ln n - ln ln n +
\Theta(1)$~\cite{Set-Cover-s,Set-Cover-l,Set-Cover-sr}. By~\cite{
Set-Cover-rs,Set-Cover-ams} it follows that unless ${\cal P = NP}$,
there exists a constant $0 < c < 1$ so that set cover cannot be
efficiently approximated to within any number smaller than $c \log_2
n$.

We formalize the {\em validation problem} discussed in the
introduction in the following manner: every edge in a traceroute
is an element in a universe $U$. Each traceroute is modeled as
a set of elements in $U$ - its edges. Each agent is modeled as a
family of sets, indicating the list of traceroutes it can perform.
Moreover, each agent has a weight, indicating the number of
traceroutes it can perform at a time period. Thus we get the
following problem:
\begin{prob}{\em Validation Set Cover - VSC}
Given a universe $U$ of $n$ elements, a collection of subsets of
$U$, ${\cal S} = \{ S_{1},...,S_k \}$, a partition of $\cal S$
$\pi=\{A_1,...,A_m\}$ where $A_i\subseteq \cal S$, and a weight
function $\omega: \pi \rightarrow \mathbb{N}$, find a subcollection
$\bar{\cal S}$ of $\cal S$ that covers all elements of $U$ such that
$\max_{1\le i\le m}{\left\lceil{\frac{|A_i\cap \bar{\cal
S}|}{\omega(A_i)}}\right\rceil}$ is minimum.
\end{prob}

Note: the Validation Set Cover problem is indeed a generalization of
the set cover problem -- if $m=1$ % and $\omega(A_i)=1$ for all $1\lei\le m$
then the Validation Set Cover problem is exactly the set
cover problem. Thus the Validation Set Cover problem is also ${\cal
NP}$-hard.

\section{An $O(\log n)$-Approximation Algorithm}\label{approx.sec}
In this section we give an approximation algorithm for the VSC
problem with an approximation ratio of $O(\log n)$. We then show
that this is the best ratio possible by showing a lower bound of
$\Omega(\log n)$ on the approximation ratio.

The greedy strategy applies naturally to the VSC problem:
iteratively for each $1\le i\le m$ pick $\omega(A_i)$ sets in $A_i$
that cover the maximum number of elements in $U$ that are still
uncovered. The algorithm stops when all the elements in $U$ are
covered, and outputs the number of steps preformed.

\BA\label{al:greedy-gsc}{\em Greedy VSC algorithm} \BE
\item $\ell\leftarrow 0$ \item $C\leftarrow\phi$ \item while $C\neq U$
\BE \item $\ell\leftarrow \ell+1$ \item for $1\le i \le m$ \BE
\item repeat $\omega(A_i)$ times\BE\item find a set $S_j$ such that
$S_j\in A_i$ and $S_j\cap(U\setminus C)$ is maximum.
\item pick $S_j$ \item $C\leftarrow C\cup S_j$\EE\EE\EE \item output
$\ell$\EE \EA

\BT\label{theo:main} Algorithm~\ref{al:greedy-gsc} gives an
approximation ratio of $O(\log n)$.\ET

We next prove Theorem~\ref{theo:main}. We first define the
$\ell$-residual VSC problem. The input to this problem is the input
to the VSC problems after $\ell$ steps of the algorithm, with the
same objective function: \BI
\item Let $n_{\ell}$ be the number of elements in $U$ that remain after
$\ell$ steps of the algorithm. For $\ell=0$ $n_{\ell}=n$.
%\item let $OPT$ be the optimal solution, let $OPT_{\ell}$ be the
%optimal solution of the residual input after $\ell$ steps.
\item let $C_{\ell}$ be the set of
elements in $U$ that are covered until step $\ell$,
\item for
all $1\le j\le k=|{\cal S}|$ \BI
\item
%such that $S_j$ hasn't been picked by step $\ell$,
let $S^{\ell}_j = S_j\setminus C_{\ell}$, %where $C_{\ell}$ is the
%set of elements that are covered until step $\ell$,
\item for all $1\le i\le m$ let $A^{\ell}_i = A_i\setminus\{S_j\in A_i| S_j$ has been
picked until step $\ell\}$, \item let ${\cal S}^{\ell} =
\{S^{\ell}_j|S^{\ell}_j\neq \phi\}$. \EI \item for all $1\le i\le m$
let $\omega(A^{\ell}_i)=\omega(A_i)$. \item let
$\textsc{opt}_{\ell}$ be the optimal solution of the residual input
after $\ell$ steps.\footnote{Recall that $\textsc{opt}$ is the
optimal solution} \EI Then $\textsc{opt}_{\ell} = \min_{\bar {\cal
S}^{\ell}}{\max_{1\le i\le m}{\left\lceil{\frac{|A^{\ell}_i\cap
\bar{\cal S}^{\ell}|}{\omega(A^{\ell}_i)}}\right\rceil}}$ where
$\bar {\cal S}^{\ell}$ is a subcollection of $\cal S^{\ell}$ that
covers all elements of $U\setminus C_{\ell}$.

Thus we get the following claim: \BCM At step $\ell\ge 1$ of
Algorithm~\ref{al:greedy-gsc} at least
$\frac{n_{\ell-1}}{\textsc{opt}_{\ell-1}}$ elements in $U$ are
covered.\ECM \BPF If $\ell=1$ then, since
Algorithm~\ref{al:greedy-gsc} picks a set that covers the maximum
number of elements, it holds that at least
$\frac{n}{\textsc{opt}}=\frac{n_{\ell-1}}{\textsc{opt}_{\ell-1}}$
elements are covered at step $\ell$. If $\ell>1$ then an optimal
algorithm covers all the $n_{\ell-1}$ remaining elements of
$U\setminus C_{\ell-1}$ in $\textsc{opt}_{\ell-1}$ steps. Since
Algorithm~\ref{al:greedy-gsc} picks a set that covers the maximum
number of remaining elements, it holds that at least
$\frac{n_{\ell-1}}{\textsc{opt}_{\ell-1}}$ elements are covered at
step $\ell$. \EPF

Using the above claim and the observation that for all $\ell$
$\textsc{opt}_{\ell}\le \textsc{opt}$, we get the following lemma.
\BL $n_{\ell}\le n\left(1-\frac{1}{\textsc{opt}}\right)^{\ell-1}$\EL
\BPF By induction on $\ell$: \[n_1\le n-\frac{n}{\textsc{opt}} =
n\left(1-\frac{1}{\textsc{opt}}\right)\]
\[n_2\le n_1-\frac{n_1}{\textsc{opt}_1}\le n\left(1-\frac{1}{\textsc{opt}}\right) -
\frac{n_1}{\textsc{opt}_1}\le n\left(1-\frac{1}{\textsc{opt}}\right)
- \frac{n\left(1-\frac{1}{\textsc{opt}}\right)}{\textsc{opt}} =
n\left(1-\frac{1}{\textsc{opt}}\right)^2
\] Assume that for all $i< \ell$ it holds that
$n_i\le n\left(1-\frac{1}{\textsc{opt}}\right)^{i}$.
Then \BEQN n_{\ell}&\le& n_{\ell-1} -
\frac{n_{\ell-1}}{\textsc{opt}_{\ell-1}}\le
n\left(1-\frac{1}{\textsc{opt}}\right)^{\ell-1} -
\frac{n_{\ell-1}}{\textsc{opt}_{\ell-1}}\nonumber\\&\le&
n\left(1-\frac{1}{\textsc{opt}}\right)^{\ell-1} -
\frac{n\left(1-\frac{1}{\textsc{opt}}\right)^{\ell-1}}{\textsc{opt}}
= n\left(1-\frac{1}{\textsc{opt}}\right)^{\ell} \EEQN\EPF

\BPFOF{Theorem~\ref{theo:main}} In the worst case the algorithm
stops after $\ell+1$ steps for the minimal $\ell$ such that
$n_{\ell}\le 1$. Since by the above lemma $n_{\ell}\le
n\left(1-\frac{1}{\textsc{opt}}\right)^{\ell}$, for $\ell$ for which
$n\left(1-\frac{1}{\textsc{opt}}\right)^{\ell}\le 1$ it holds that
$n_{\ell}\le 1$.

\BEQN n\left(1-\frac{1}{\textsc{opt}}\right)^{\ell}&\le& 1
\Leftrightarrow \left(1-\frac{1}{\textsc{opt}}\right)^{\ell}\le
\frac{1}{n}\nonumber\\&\Leftrightarrow& \ell\le
\frac{\log(1/n)}{\log \left(1-\frac{1}{\textsc{opt}}\right)} =
\frac{\log n}{\log
\left(\frac{\textsc{opt}}{\textsc{opt}-1}\right)}\nonumber\\&\Leftrightarrow&
\ell\le \frac{\log n}{\log
\left(1+\frac{1}{\textsc{opt}-1}\right)}.\EEQN We now prove that
$\frac{\log n}{\log \left(1+\frac{1}{\textsc{opt}-1}\right)}\le \log
n \cdot \textsc{opt}$. It holds that
\[\frac{\log n}{\log \left(1+\frac{1}{\textsc{opt}-1}\right)}\le \log n \cdot
\textsc{opt} \Leftrightarrow 1+\frac{1}{\textsc{opt}-1}\ge
e^{1/\textsc{opt}}.\]
 According to Taylor series have that
\[f(x)=\sum_{i=0}^n{f^{(i)}(0)\frac{x^i}{i!}}+R_n(x),\] where
\[R_n(x) = \frac{f^{(n+1)}(c)}{(n+1)!}x^{n+1},\] for some $0\le c\le x$.
For $f(x)=e^x$ we get that
\[e^x=\sum_{i=0}^n{\frac{x^i}{i!}}+ e^c\frac{x^{n+1}}{(n+1)!},\] for some $0\le c\le x$.
 For $x=1/\textsc{opt}$ and $n=2$ we get that \[e^{1/\textsc{opt}}=
 1+\frac{1}{\textsc{opt}}+\frac{1}{2\textsc{opt}^2}+\frac{e^c}{6\textsc{opt}^3},\]
 for some $0\le c\le 1/\textsc{opt}$. Now, \BEQN 1+\frac{1}{\textsc{opt}-1}&\ge&
 1+\frac{1}{\textsc{opt}}+\frac{1}{2\textsc{opt}^2}+\frac{e^c}{6\textsc{opt}^3}\nonumber\\
&\Leftrightarrow& \frac{1}{\textsc{opt}-1}-\frac{1}{\textsc{opt}}\ge
\frac{1}{2\textsc{opt}^2}+\frac{e^c}{6\textsc{opt}^3}\nonumber\\
 &\Leftrightarrow& \frac{1}{(\textsc{opt}-1)\textsc{opt}}\ge
 \frac{1}{2\textsc{opt}^2}+\frac{e^c}{6\textsc{opt}^3}\nonumber\\
 &\Leftrightarrow& \frac{1}{\textsc{opt}-1}\ge
 \frac{1}{2\textsc{opt}}+\frac{e^c}{6\textsc{opt}^2}\nonumber\\
 &\Leftrightarrow& 6\textsc{opt}^2\ge (\textsc{opt}-1)(3\textsc{opt}+e^c).\EEQN
 The last inequality is valid since $e^c<3$ (as $c\le 1/\textsc{opt}$).
  Thus $1+\frac{1}{\textsc{opt}-1}\ge e^{1/\textsc{opt}}$, so $\frac{\log n}{\log
\left(1+\frac{1}{\textsc{opt}-1}\right)}\le \log n \cdot
\textsc{opt}$. Therefore the number of steps used by
Algorithm~\ref{al:greedy-gsc} is at most $1+\log n \cdot
\textsc{opt}$, and the theorem follows.
 \EPFOF

By~\cite{Set-Cover-rs,Set-Cover-ams} it follows that unless ${\cal
P = NP}$ the approximation ratio of the set cover problem is
$\Omega(\log n)$. Since for $m=1$ and for $1\le i\le m$
$\omega(A_i)=1$ the VSC problem is exactly the set cover problem, we
get that unless ${\cal P = NP}$ the approximation ratio of the VSC
problem is $\Omega(\log n)$.

\end{document}